\documentclass[twocolumn,preprintnumbers,amsmath,amssymb,pre]{revtex4}

\usepackage{txfonts}
\usepackage{amsfonts}
\usepackage{amsmath}
\usepackage{graphicx}
\usepackage{dcolumn}
\usepackage{bm}
\usepackage{overpic}
\usepackage{multirow}

\begin{document}

\title{Effect of an interstitial fluid on the dynamics of three-dimensional granular media}

\author{Rene Zu\~{n}iga$^{1,2}$, St{\'e}phane Job$^{2}$ and Francisco Santibanez$^{1,3}$}

\email[Corresponding author: ]{francisco.santibanez@unc.edu}

\affiliation{$^{1}$ Instituto de F\'isica, Pontificia Universidad Cat\'olica de Valpara\'iso,\\ Av. Brasil 2950, Valpara\'iso, Chile.\\$^{2}$ Laboratoire Quartz, EA 7393, Supm{\'e}ca, 3 rue Fernand Hainaut 93400 Saint-Ouen, France.\\ $^{3}$ Joint Department of Biomedical Engineering, University of North Carolina at Chapel Hill and the North Carolina State University, Chapel Hill, North Carolina, USA.}

\date{\today}

\begin{abstract}
The propagation of mechanical energy in granular materials has been intensively studied in recent years given the wide range of fields that have processes related to this phenomena, from geology to impact mitigation and protection of buildings and structures. In this paper, we experimentally explore the effect of an interstitial fluid on the dynamics of the propagation of a mechanical pulse in a granular packing under controlled confinement pressure. The experimental results reveal the occurrence of an elastohydrodynamic mechanism at the scale of the contacts between wet particles. We describe our results in terms of an effective medium theory, including the presence of the viscous fluid. Finally, we study the nonlinear weakening of the granular packing as a function of the amplitude of the pulses. Our observations demonstrate that the softening of the material can be impeded by adjusting the viscosity of the interstitial fluid above a threshold at which the elastohydrodynamic interaction overcomes the elastic repulsion due to the confinement.
\end{abstract}

\maketitle

\section{\label{sec:intro}Introduction}

The propagation of mechanical vibrations in particulate matter is a nontrivial phenomenon that has implications in several fields, both in basic and applied research~\cite{deGennes1999,Porter2015}. The complexity of the contact network~\cite{Radjai1996,Lherminier2014} and the nonlinearity of the interaction force between particles~\cite{Popov2010} determines how the waves propagate in granular media~\cite{Norris1997}. In dry packings, the repulsive nonlinear interaction between two spheres with radius $R$ and elastic modulus $E$ relies on the Hertz contact force, $F_H \propto ER^{1/2}\delta^{3/2}$~\cite{Landau1986}. The latter implies that the contact stiffness increases rapidly with the deformation $\delta$ and vanishes in absence of mechanical contact. In the dry case, there is thus no tensile force, and the particles can eventually loose contact~\cite{Tournat2004}. In one-dimensional lattices of particles~\cite{Nesterenko2001}, the nonlinear elastic interaction given by the Hertzian contact dictates the dynamics; a mechanical impact can propagate weakly to strongly nonlinear waves~\cite{Daraio2005,Job2005} depending on the amplitude of the pulse. Similarly, in random granular packing, a low-amplitude mechanical excitation generates a linear ballistic waves, both in the longitudinal (namely a {\em P-wave}) and in the transverse (namely a {\em S-wave}) directions~\cite{Norris1997,Makse2004,Johnson2005}. The coherent perturbation, resulting from an ensemble average~\cite{Page1995}, travels straight from the source to the receiver, as in an effective medium~\cite{Norris1997,Makse2004}. Owing to the randomness, the coherent pulse is followed by an incoherent and long-lasting {\em coda wave}, which corresponds to the multiple scattering~\cite{Snieder2006} of the initial excitation, across the contact network. Interestingly, the ballistic waves possess the reminiscent features of the microscopic scale~\cite{Makse2004}. Recently, a continuous description bridging the linear and the nonlinear regimes in random packing of particles has been proposed~\cite{Wildenberg2013}, depending on the confinement pressure applied to the packing and the amplitude of the propagating impulse~\cite{Santibanez2016}.\\

Weakening of granular materials is another nonlinear mechanism that is reported to occur in particulate systems. Its origin is a softening induced by an acoustic fluidization mechanism~\cite{Jia2011, Espindola2012, Giacco2015, Olson2015}. Such nonlinear process has been demonstrated to be responsible for triggering secondary earthquakes after the occurrence of a main seismic event~\cite{Johnson2005}. The propagation of the mechanical impulse interacts with the grains in the material and mobilizes the particles with weaker contacts~\cite{Johnson2008, Johnson2016}. This leads to a modification of the contact network that can trigger major faults and lead to the emission of new events~\cite{Johnson2005,Johnson2008}. Since the material weakening involves weakly consolidated contacts, it can be modified by the presence of an interstitial fluid, owing to an enhanced cohesion due to capillary effects~\cite{Dorostkar2018} or to the viscous lubrication between the grains~\cite{Galaz2018}.\\

Wet granular media are ubiquitous in nature, as for instance the sediments, the mud, or the sand of a beach. In this paper, we look at understanding how the presence of an interstitial fluid can affect the nonlinear dynamics of particulate matter. The behavior of wet particles has been investigated in recent years due to the great number of industrial applications, from mining to food and pharmaceutical industries. A wet granular medium has a fluid phase that partially occupies the interstitial volume available between grains and their motions are likely determined by interactions mediated by the fluid. In the literature, it has been shown that viscous forces, surface tension, and capillary bridges among others, have a great importance in systems where the energy dissipation occurs due to liquid films trapped either in the asperities~\cite{Brunet2008} or nearby the contact region~\cite{Marshall2011}. On one hand, capillarity bridges between grains~\cite{Semprebon2016} have been studied owing to significant effects on the static cohesion of granular packings~\cite{Pacheco2012,Saingier2017}. On the other hand, wave propagation in wet granular media relies on the dynamics of the interstitial fluid~\cite{Herminghaus2005,Moller2007}, which generates a viscous repulsion~\cite{Donahue2010,Arutkin2017} that generally cannot be neglected at acoustic frequencies; the fluid modifies the acoustic features due to a velocity-dependent viscous contribution~\cite{Herbold2006,Job2008,Griffiths2010}, in addition to increasing the overall dissipation~\cite{Brunet2008}. When the fluid is highly confined, in the lubrication regime~\cite{Davis1986}, and under high-frequency or high-amplitude vibrations, the viscous forces can ultimately induce elastic deformations of the particles, via an elastohydrodynamic interaction~\cite{Leroy2011,Leroy2012,Villey2013,Wang2015}. Understanding all these behaviors is fundamental for unraveling all the phenomena ranging from the stability of a pile of wet grains~\cite{Scheel2008} to the dynamics of dense suspensions~\cite{Waitukaitis2012,Han2016,Peters2016,Buttinoni2017}.\\

In this paper, we aim (i) at ruling out the mechanisms involved in the propagation of mechanical impulses in a wet model granular medium and (ii) at unraveling how the fluid can affect the dynamic weakening of such a material. In Sec.~\ref{sec:setup_observations}, we present the experimental setup, the protocols of analysis, and a description of the experimental observations. In Sec.~\ref{sec:model}, we analyze and interpret the features of the high-frequency spectrum of the transmitted pulses, in terms of the propagation velocity. These features allow probing the elasticity of the medium, and how it is affected by the presence of an interstitial fluid. In Sec.~\ref{sec:weakening}, we analyze the low-frequency spectrum of the mechanical response. Our measurements reveal the elastic weakening of dry granular media at low confinement and high amplitude, which disappears either at large confinement pressure or when the contacts are lubricated by a sufficiently viscous fluid. Finally, Sec.~\ref{sec:conclusion} summarizes the results and the observations presented in this paper.\\

\section{\label{sec:setup_observations} Experimental Setup and observations}

\begin{figure}[b]
\begin{center}
\includegraphics[width=0.450\textwidth]{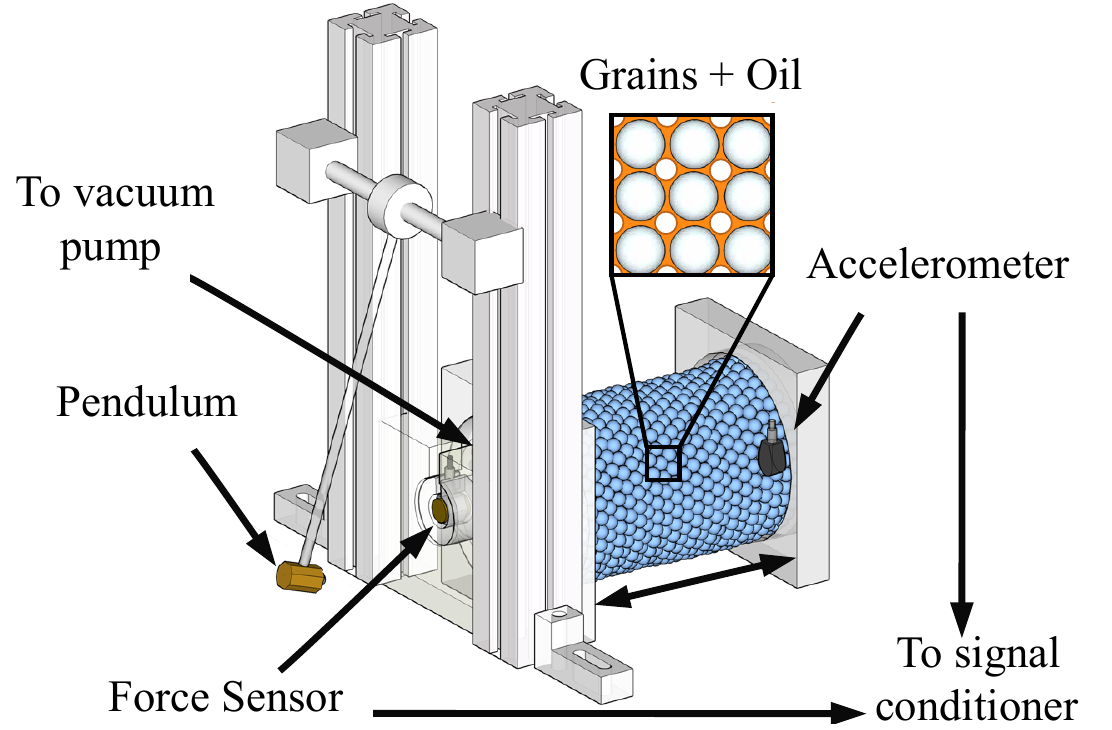}
\caption{\label{fig:setup} Sketch of the experimental setup depicting a cylindrical granular medium enclosed in a soft elastic sheet, the shock initiator pendulum, and the positions of the dynamic sensors.}
\end{center}
\end{figure}

The experimental sample under study (see Fig.~\ref{fig:setup} and Ref.~\cite{Santibanez2016}) consists in a packing of approximately $3000$ spherical glass particles (density $\rho_g=2400$~kg/m$^3$, radius $R_g=2.5$~mm, Young modulus $E_g=69$~GPa, Poisson's ratio $\nu_g=0.2$, and surface roughness $Ra\simeq10$~nm~\cite{Brunet2008,Buttinoni2017}) confined inside a thin deformable latex sheet. The sheet is hermetically sealed and clamped between two square plates made of a rigid plastic. The two plates are $1$~cm thick and have a $5$~cm in diameter circular aperture. Thin lateral holes on the side of the plates allow the emergence of sensors cables and a vacuum hose. The soft sheet allows maintaining a controlled isotropic stress on the granular medium, by evacuating the interstitial air from the hose with a vacuum pump. While the sample is set under pressure, it can be molded by hand in the form of a cylinder of length $L_s=15$~cm and radius $R_s=2.5$~cm that fits in the holes of the two supporting plastic plates. This finally leads to a solidlike granular medium with compactness approximately equal to the random close packing fraction, $\phi_s\simeq0.63$. The pump allows reaching a hydrostatic pressure of as high as $P_0\approx83$~kPa, which is probed by a static pressure sensor ({\em Honeywell 19C015PV5K} with its {\em INA114} low-noise amplifier). At one extremity of the sample, a dynamic force sensor ({\em PCB Piezotronics 208C01}) is placed in direct contact with the grains through a central and hermetic hole cut in the sheet. At the opposite side of the sample, a miniature accelerometer ({\em PCB Piezotronics 352A24}) is located on the axis of the cylindrical sample. A single short mechanical impulse is initiated in the sample by impacting the back of the dynamic force sensor with a $21$~cm long pendulum. The impacting head of the pendulum consists in a piece of brass in front of which a spherical glass particles has been glued. The contact duration depends on the mass ($57$~g) of the head~\cite{Landau1986}, which has been chosen so that the initial excitation is about a fraction of a millisecond~\cite{Santibanez2016}. A collision thus generates a broadband excitation from dc to few kilohertz. The strength of the impact is controlled by adjusting the initial release angle of the pendulum. At the opposite end, the accelerometer records the outgoing pulse that travels through the sample. The signals of the dynamic force sensor and the accelerometer are routed through a signal conditioner ({\em PCB Piezotronics model 482C}) and acquired simultaneously, together with the signal of the static pressure, with an analog-to-digital converter ({\em National Instruments USB-6356}) at $500$~kS/s sample rate.\\

\begin{figure*}
\includegraphics[width=0.900\textwidth]{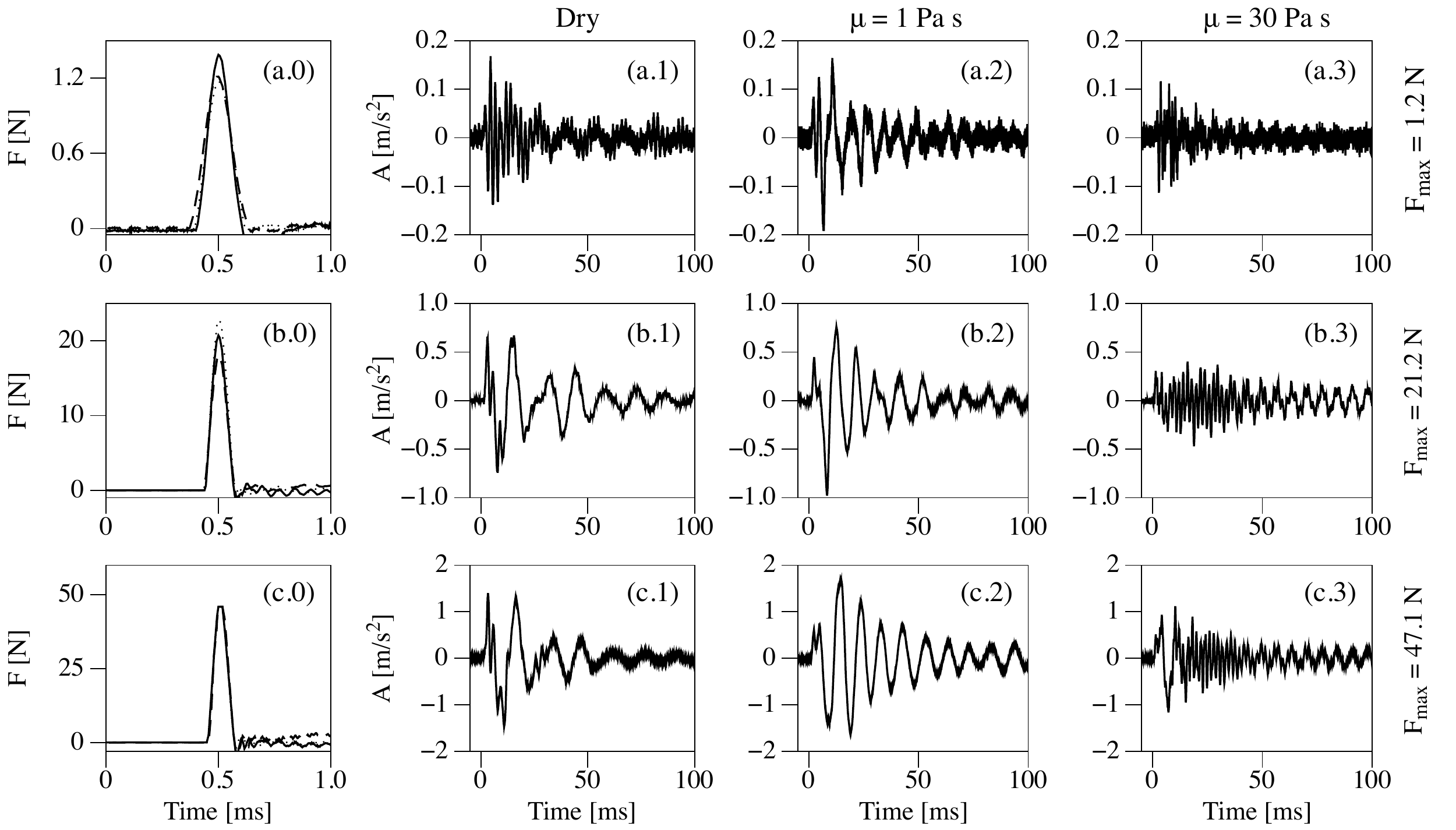}
\caption{\label{fig:signals_low_P} Examples of experimental data at low confinement pressure ($P_0=3.2$~kPa). The left column (0) shows the input force for each experiment: the solid line corresponds to the dry configuration, the dashed line is $\mu=1$~Pa.s and the dotted line is $\mu=30$~Pa.s, whose acceleration outputs are shown in the next three columns (1)--(3), respectively. Rows (a)--(c) correspond to different strength of the initial excitation, as shown in the right-side text labels.}
\end{figure*}

\begin{figure*}
\includegraphics[width=0.900\textwidth]{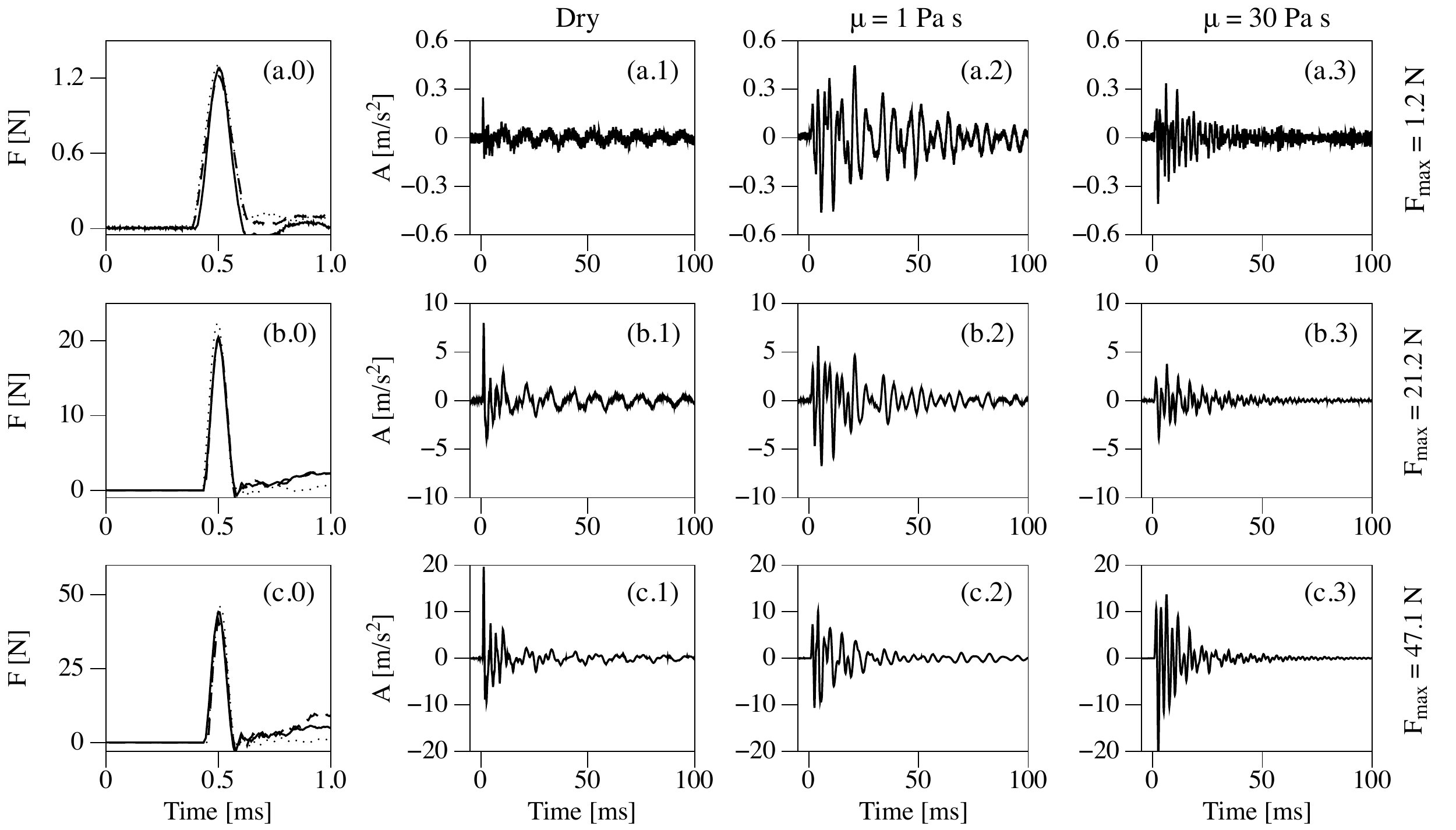}
\caption{\label{fig:signals_high_P} Examples of experimental data at high confinement pressure ($P_0=83$~kPa). The left column (0) shows the input force for each experiment: the solid line corresponds to the dry configuration, the dashed line is $\mu=1$~Pa.s and the dotted line is $\mu=30$~Pa.s, whose acceleration outputs are shown in the next three columns (1)--(3), respectively. Rows (a)--(c) correspond to different strength of the initial excitation, as shown in the right-side text labels.}
\end{figure*}

The experimental setup allows probing the acoustic response of dry and wet granular media by measuring the speed of the waves that propagate through the samples. The wet samples are prepared by homogeneously filling a fraction of the interparticle space with a viscous fluid. The amount of added liquid corresponds to a third of the available volume; given the estimated packing fraction $\phi_s$, the liquid thus approximately occupies $12\%$ of the total volume of a sample. If all the spherical particles were homogeneously lubricated~\cite{Galaz2018}, the liquid coating would have a thickness $D_{coat}=R_g([1+(1-\phi_s)/3\phi_s]^{1/3}-1)\simeq 150$~$\mu$m. Before pouring the wet particles into the soft elastic sheet, we make sure that all the grains are uniformly coated with fluid by carefully stirring the mixture in a container. After having used a fluid, the grains are washed several times with alcohol and then gently dried in an oven. The fluids are silicone oils from {\em Sigma Aldrich}. Four different fluids' viscosity were considered in this study, ranging over more than two decades, from $\mu=0.1$~Pa.s to $\mu=30$~Pa.s at room temperature. The sound speed and the mass density of these fluids are close to the features of water, $c_f\simeq1500$~m/s and $\rho_f\simeq1000$~kg/m$^3$. Their surface tension is $\gamma_f\simeq21.5$~mN/m~\cite{Langley2017} and their storage modulus is $G\simeq5$~kPa~\cite{Oswald2014}, both independently of the viscosity. In particular, these fluids remain Newtonian below a critical flow's shear rate $\dot{\gamma}_f=G/\mu$~\cite{Oswald2014,Guyon2001}. This condition is always satisfied in all our experiments, as demonstrated in Sec.~\ref{sec:model}. Above the critical shear rate, the behavior of these fluids is known to be relatively independent of the viscosity~\cite{Oswald2014,Barlow1964}.\\

\begin{figure}[b]
\includegraphics[width=0.450\textwidth]{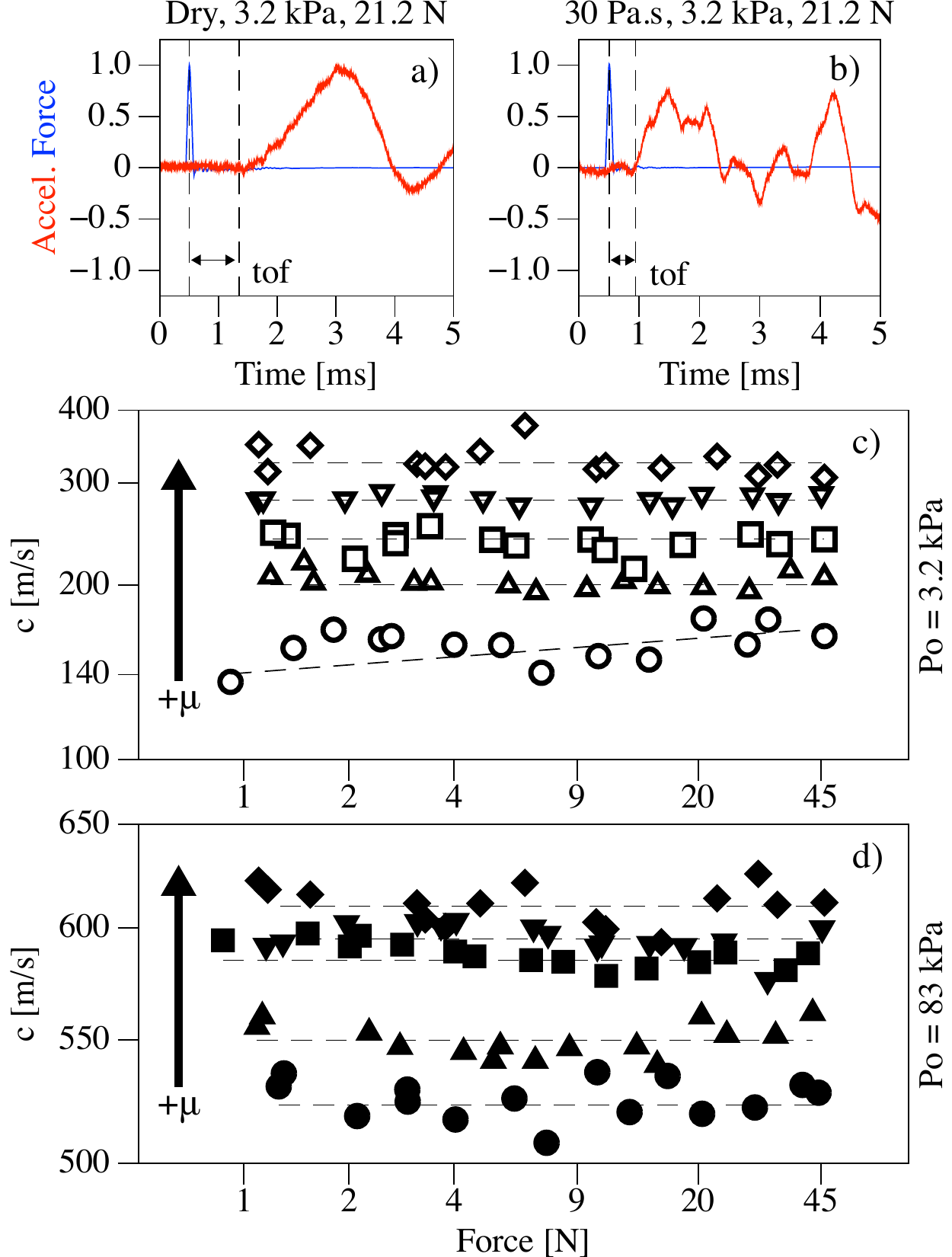}
\caption{\label{fig:wave_speed} Examples of the time of flight (tof) measurement for (a) dry and (b) wet ($\mu=30$~Pa.s) cases, at low confinement pressure. Wave speed as a function of the magnitude of the initial collision force in dry and wet granular media, for both low (c) and high (d) confinement pressures. Markers correspond to different interstitial fluid viscosity: $\circ:$ dry, $\triangle:0.1$~Pa.s, $\square:1$~Pa.s, $\triangledown:10$~Pa.s, and $\diamond:30$~Pa.s. The dashed lines are guides for the eyes; the vertical arrows point toward increasing viscosities.}
\end{figure}

A set of typical waveforms propagated through different samples under $P_0=3.2$~kPa and $P_0=83.0$~kPa confinement pressures is shown in Figs.~\ref{fig:signals_low_P} and~\ref{fig:signals_high_P}, respectively. In both figures, the rows stand for three different excitation amplitude. The leftmost columns show the initial force perturbation as a function of time, resulting from the impact of the pendulum. The short initial impact is about $0.2$~ms wide: its spectrum (not shown; see Ref.~\cite{Santibanez2016}) extends from dc to $8$~kHz approximately. The three rightmost columns present the acceleration of the transmitted waves, first in a dry medium and then with two different interstitial viscosities. The long-lasting transmitted {\em coda wave} corresponds to the multiple scattering of the incident pulse~\cite{Snieder2006}, along different paths through the random network of particles. The very first transmitted event [see the magnified plots in Figs.~\ref{fig:wave_speed}(a) and~\ref{fig:wave_speed}(b)] corresponds to the fastest ballistic contribution~\cite{Page1995}, that is a longitudinal pressure wave traveling straight from the source to the receiver~\cite{Norris1997, Makse2004, Johnson2005}. In all the transmitted signals presented in Figs.~\ref{fig:signals_low_P} and~\ref{fig:signals_high_P}, one clearly distinguishes two distinct frequency components. On one hand, the ballistic wave carries a high-frequency content, lying in the range of $1$~kHz [see the rise time in Figs.~\ref{fig:wave_speed}(a) and~\ref{fig:wave_speed}(b)]. This spectrum matches the bandwidth of the excitation attenuated by the frequency-dependent scattering~\cite{Page1995} and the viscoelastic dissipation at the contact between grains~\cite{Job2005, Popov2010}. In the presence of a fluid, the amplitude of the ballistic pulse decreases slightly more: the viscous dissipation enhances the attenuation~\cite{Herbold2006,Job2008,Brunet2008}. The analysis of the speed of the ballistic wave is presented in more details in Sec.~\ref{sec:model}. On the other hand, the transmitted waves shown in Figs.~\ref{fig:signals_low_P} and~\ref{fig:signals_high_P} also exhibit a long-lasting, superimposed, low-frequency oscillation of the order of $100$~Hz. It has been demonstrated to rely on a longitudinal resonance of the sample~\cite{Santibanez2016} and to reveal the nonlinear weakening~\cite{Jia2011, Espindola2012, Olson2015, Giacco2015, Johnson2008, Johnson2016} of the granular matter. The detailed analysis of the low-frequency oscillation is presented in Sec.~\ref{sec:weakening}.\\

The measurements of the speed of the ballistic pulses are presented in Fig.~\ref{fig:wave_speed}. The wave speed $c$ is estimated as the distance $L_s$ between the input (force sensor) and the output (accelerometer) divided by the time of flight. The latter is measured as the time difference between the earliest in and out events [see for instance the dashed line in Figs.~\ref{fig:wave_speed}(a) and~\ref{fig:wave_speed}(b)], which are detected systematically as the instants at which the amplitude of each waveform emerges above a threshold (three times the average noise level). In Figs.~\ref{fig:wave_speed}(c) and~\ref{fig:wave_speed}(d), we show the propagation velocity for both low and high confinement pressures, respectively. In both cases, the wave speed is measured first in a dry medium as a reference, then with the four fluids with different viscosities. The dry measurements at low confinement pressure, $P_0=3.2$~kPa, are shown as circles in Fig.~\ref{fig:wave_speed}(c): the log-log representation reveals that the wave speed $c$ approximately increases as a power law of the magnitude of the excitation, as it is predicted from a Hertzian interaction~\cite{Norris1997, Makse2004, Santibanez2016, Wildenberg2013}, see Sec.~\ref{sec:model}. In contrast, with interstitial fluids, the wave speed increases more slowly with the strength of the impact and quickly tends to a constant value, independent of the input force, at large viscosities. Remarkably, the propagation velocity noticeably increases with the viscosity of the fluid; this nontrivial feature has been observed yet in one-dimensional granular media~\cite{Herbold2006,Job2008} and is analyzed in Sec.~\ref{sec:model}. On the opposite, at the highest confinement pressure, $P_0=83$~kPa, the propagation velocity shown in Fig.~\ref{fig:wave_speed}(d) appears independent on the amplitude of the perturbation in the dry configuration: the dynamical stress being negligible compared to the static pressure, the response is linear. With fluid, the wave speed does not depend either on the perturbation strength and a weaker dependence of the wave speed with the viscosity of the fluid is observed, compared to the low confinement case.\\

These observations suggest a competition between the elastic deformations of the particles resulting from the confinement or the dynamic pressure, and an effect of the viscous fluid, which likely resides in between the particles and at the periphery of their contacts~\cite{Marshall2011}. Here, it is possible to rule out the contribution of capillary effects since all our fluids have the same surface tension, $\gamma_f\simeq21.5$~mN/m, independently of their viscosity~\cite{Langley2017}: it thus cannot explain the viscosity-dependent wave speed observed in the wet media. Moreover, equating the Laplace pressure~\cite{Guyon2001} to the confinement pressure, $P_0\simeq \gamma_f/R_c$, indicates excessively small capillary bridges curvatures, compared to the size of the particles and the liquid's filling fraction, to produce a significant contribution: $R_c\simeq P_0/\gamma_f\sim 7$~$\mu$m at $P_0=3.2$~kPa and $R_c\sim 0.25$~$\mu$m at $P_0=83$~kPa. The next section aims at deriving an alternative framework, in order to relate the increase of the wave speed, i.e., the enhancement of the effective elasticity of the medium, as an effect of the viscous flow in the interstices of the particles.\\

\section{\label{sec:model} Wave speed analysis and interpretation}

\subsection{\label{subsec:model_dry} Wave speed in dry granular media}

Measuring the speed of mechanical waves gives access to the elastic features of the media, at an effective macroscopic scale. Considering that the fastest event corresponds to a ballistic pressure wave in the longitudinal direction~\cite{Norris1997, Makse2004}, one can extract the effective longitudinal modulus $M$ from the experimental measurements of the wave speed $c$ shown in Fig.~\ref{fig:wave_speed},
\begin{equation}
M = K+(4/3)G = \rho_g\phi_s c^2.\label{eq:model_Mexp}
\end{equation}

According to the effective medium theory (EMT)~\cite{Norris1997, Makse2004}, the effective bulk and shear moduli, $K$ and $G$, of a random packing of frictional spheres can be related to the interactions at the interparticles scale; these moduli are given by
\begin{eqnarray}
K &\propto& (Z\phi_s/R_{\ast})\kappa_n,\label{eq:model_Kemt}\\
G &\propto& (Z\phi_s/R_{\ast})[\kappa_n+(3/2)\kappa_t],\label{eq:model_Gemt}
\end{eqnarray}
where $\kappa_{n,t}=\partial F_{n,t}/\partial \delta_{n,t}$, $F_{n,t}$, and $\delta_{n,t}$ are the normal and tangential stiffnesses, forces, and deformations at a single contact between two particles, respectively. $Z$ is the coordination number and $R_\ast=R_g/2$ is the reduced radius of curvature at the contacts. The interparticles stiffness depends on the size of the mechanical contact; in the case of dry spheres, the Hertz-Mindlin interaction potential~\cite{Landau1986, Popov2010} provides
\begin{eqnarray}
a_{dry} &=& (R_\ast\delta_n)^{1/2}\propto R_\ast(p/E_\ast)^{1/3},\label{eq:model_Adry}\\
\kappa_{n,t} &\propto& F_{n,t}/\delta_{n,t} \propto E_\ast a_{dry},\label{eq:model_Kdry}
\end{eqnarray}
where $a_{dry}$ is the radius of the flat contact disk between two spheres, $E_{\ast}=E_g/2(1-\nu_g^2)$ is the reduced elastic modulus and $p\propto Z\phi_sF_n/R_{\ast}^2$~\cite{Norris1997} is the confining pressure. Note that the scaling given in Eq.~\ref{eq:model_Kdry} relies on the estimation of the normal force, $F_n\propto\pi a_{dry}^2p_{max}\propto E_\ast a_{dry}\delta_n$, from the maximal pressure inside the contact region given by the Hooke's law, $p_{max}\propto E_\ast(\delta_n/a_{dry})$. Note also that in Eq.~\ref{eq:model_Gemt}, the tangential contribution $\kappa_t$ stands for a nonsliding (i.e., sticking) contact resulting from an infinite Coulomb's friction coefficient between particles. In the case of frictionless particles, the tangential stiffness is zero, $\kappa_t=0$. Hence, it turns out that the effective longitudinal modulus non-linearly depends on the confinement pressure $p$,
\begin{equation}
(M_{dry}/E_{\ast}) = \alpha_{dry}(p/E_{\ast})^{1/3},\label{eq:model_Mdry}
\end{equation}
where $\alpha_{dry}$ is a numerical prefactor of the order of unity, which only depends on the topology of the granular packing (via $Z$ and $\phi_s$) and wether the contacts of the particles stick or slip. Quantitatively (see Eqs. (1), (12), (13) and (14) in Ref.~\cite{Makse2004}), a lower and an upper bounds of the prefactor are $\alpha_{dry}\simeq 0.37$ for frictionless particles with $Z=2$ and $\alpha_{dry}\simeq 1.70$ for frictional particles with $Z=6$, at $\phi_s=0.63$.\\

\begin{figure}[b]
\includegraphics[width=0.450\textwidth]{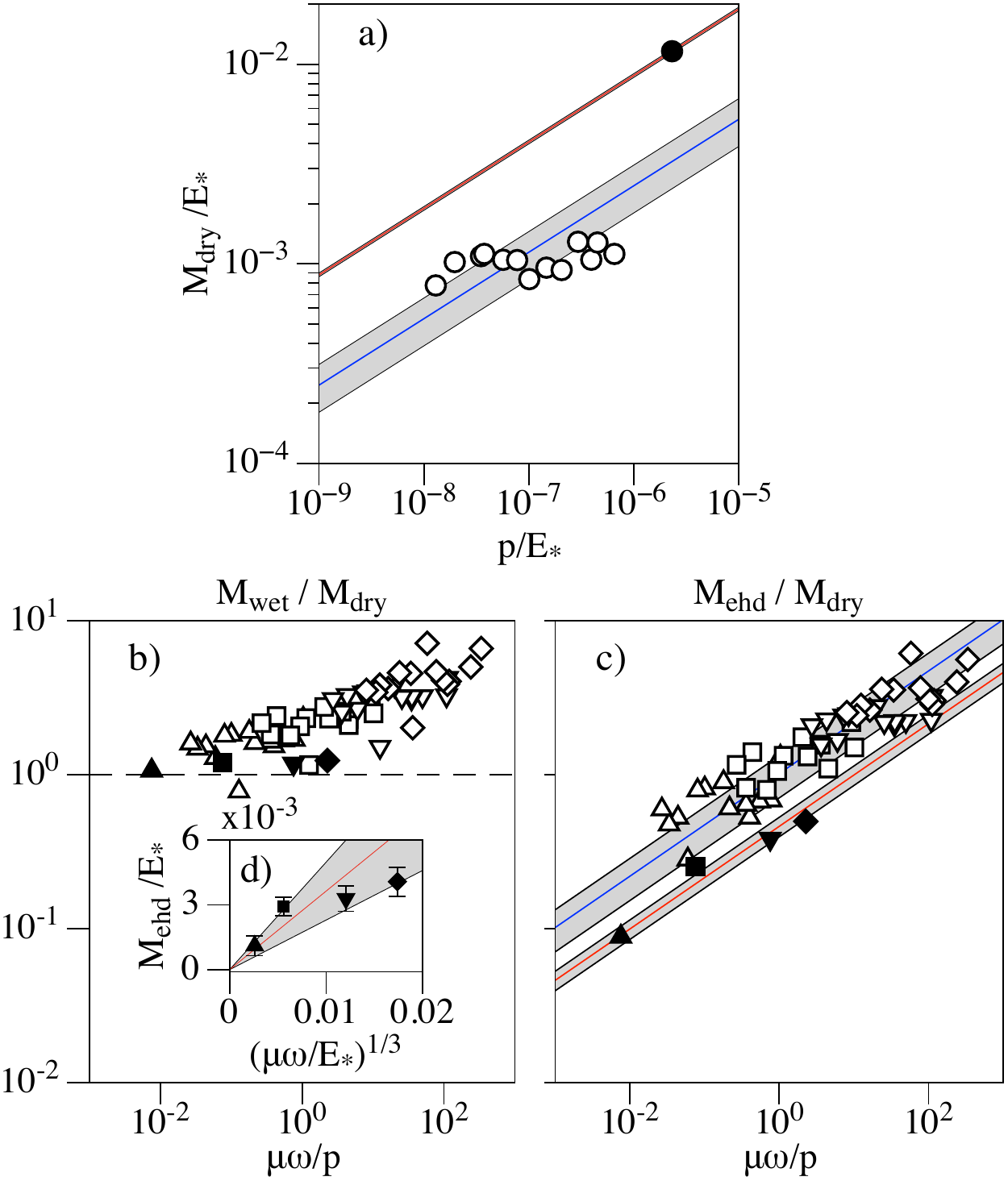}
\caption{\label{fig:elastic_moduli} (a) Effective longitudinal modulus in the dry case as a function of the pressure $p$, where $p$ stands either for the static pressure $P_0$ at high confinement or for the magnitude of the perturbation in the low confinement limit. (b) Ratio of the wet to dry longitudinal modulus, $M_{wet}/M_{dry}$, as a function of $\mu\omega/p$. (c) Ratio of the elastohydrodynamic contribution to the dry longitudinal modulus as a function of $\mu\omega/p$, with $M_{ehd}=M_{wet}-M_{dry}$, see Eq.~\ref{eq:model_Mwet}. (d) Linear plot of the elastohydrodynamic contribution to the elastic modulus as a function of $(\mu\omega/E_{\ast})^{1/3}$. In (a)--(d), the markers refer to the definition given in Fig.~\ref{fig:wave_speed} and the shaded region show the $50\%$ mean deviation error. In (a) and (c), the straight lines have a slope of $1/3$, according to Eqs.~\ref{eq:model_Mdry},~\ref{eq:model_Mehd} and~\ref{eq:model_Mwet}.}
\end{figure}

The estimations of the dry longitudinal modulus, $M_{dry}$, obtained from the measurements of the wave speeds $c$ shown in Fig.~\ref{fig:wave_speed}, are presented in Fig.~\ref{fig:elastic_moduli}(a) in a nondimensional form. Experimentally, $p$ stands for the static pressure at the highest confinement, $p=P_0+P_m\simeq P_0\gg P_m$, where $P_m=F_m/\pi R_s^2$ is an estimation of the magnitude of the dynamic stress and $F_m$ is the magnitude of the measured excitation force. At the lowest confinement, $p$ stands for the magnitude of the perturbation, $p\simeq P_m\gg P_0$. Matching the data shown in Fig.~\ref{fig:elastic_moduli}(a) to the Eq.~\ref{eq:model_Mdry} provides two estimations for the prefactor, both being of the order of unity: $\alpha_{dry}=0.25\pm38\%$ at low confinement and $\alpha_{dry}=0.88\pm3\%$ at high confinement. The order of magnitude of these coefficients are in fair agreement with the EMT prediction given in Eq.~\ref{eq:model_Mdry}. In particular, the experimental data at low confinement pressure reveal the nonlinear nature of the amplitude-dependent elastic modulus. In this regime, the slightly lower exponent, in comparison to the $1/3$ expectation, and the slightly low experimental prefactor $\alpha_{dry}$ are presumably a trace of the nonlinear softening of the material~\cite{Jia2011, Espindola2012, Giacco2015, Olson2015, Johnson2016} described in Sec.~\ref{sec:weakening}.

\subsection{\label{subsec:model_wet}elastohydrodynamic interactions mediated by the fluid}

\begin{figure}[t]
\begin{center}
\includegraphics[width=0.450\textwidth]{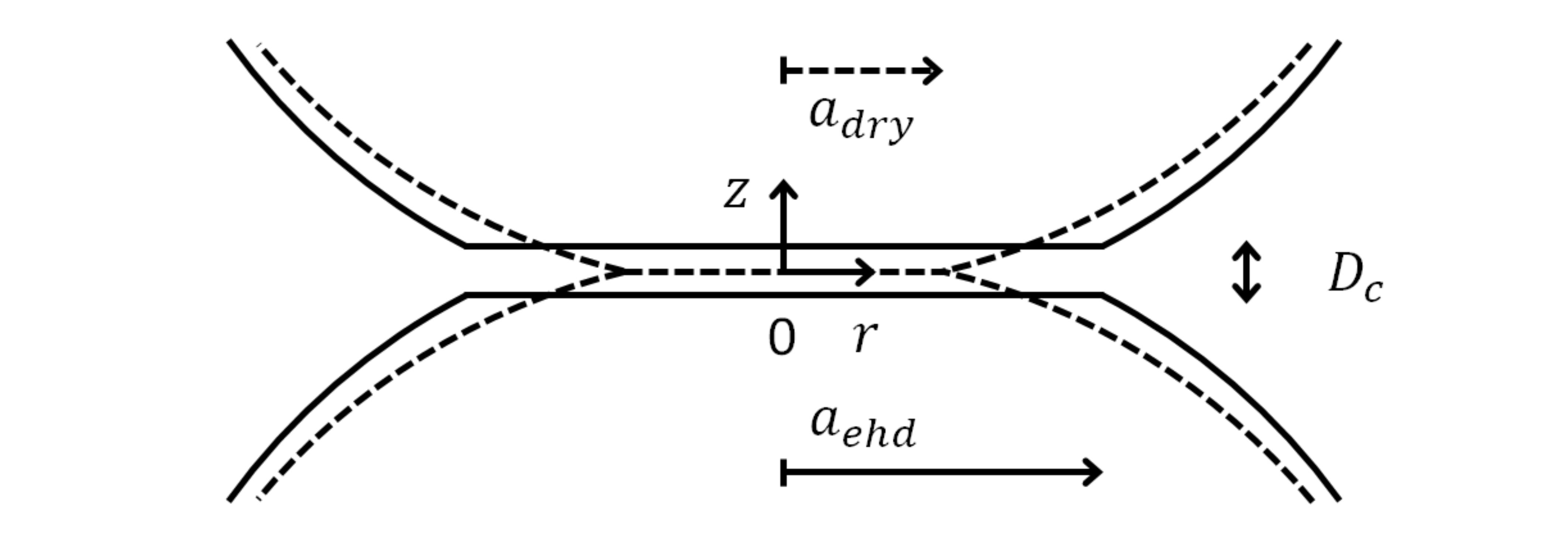}
\caption{\label{fig:sketch_contact} Sketch of the normal deformation at the contact between two spheres in the dry case (dashed line) and with an interstitial fluid (solid line), depicting the spatial extent of the deformations, $a_{dry}$ and $a_{ehd}$, and a fluid layer with thickness $D_c$.}
\end{center}
\end{figure}

Getting a deeper insight on the fluid-dependent elastic features shown in Fig.~\ref{fig:elastic_moduli} can be achieved by analyzing the interaction between two elementary elastic spheres separated by a thin layer of viscous fluid, as sketched in Fig.~\ref{fig:sketch_contact}. This case has been extensively addressed in the literature untill recently, via the elastohydrodynamic rebound of an elastic sphere on a liquid layer~\cite{Davis1986} or via the oscillatory excitations of a spherical tip immersed in a fluid near an elastic plane~\cite{Leroy2011, Leroy2012, Villey2013}. The details presented in these four studies are recalled in the following, in order to derive a useful description of our observations. In the following, $D>0$ denotes the initial separation between two spheres separated by an interstitial layer of viscous fluid in the lubrication limit, $D\ll R_\ast$. The relative displacement of the particles and their elastic deformation, mediated by the fluid, are denoted by $d(t)$ and $\delta_n(r,t)$, respectively. They result from a normal collision, along the direction $z$ in Fig.~\ref{fig:sketch_contact}, at relative velocity $v_n=\dot{d}\propto\omega d$. Here, $\omega=2\pi f$ is the angular frequency relying on the duration of the shock, $\tau\propto\omega^{-1}$. The thickness of the fluid as a function of the radial coordinate $r$ and the time $t$ thus reads $\Delta=D+\delta_n(r,t)-d(t)+r^2/2R_\ast$.\\

{\em Rigid particles ($\delta_n=0$)}. The case of rigid particles is considered first. The weak collision ($d\ll D$) between the two spheres squeezes the fluid out from the interstitial region and induces a radial flow, which is assumed laminar and incompressible~\cite{Davis1986, Guyon2001}. Within the incompressible assumption, the mean radial velocity of the fluid $\langle v_r \rangle=(1/\Delta)\int_0^\Delta{v_rdz}$ can be estimated from the flow rate conservation, $\pi r^2v_n=2\pi r\Delta\langle v_r \rangle$, as $\langle v_r \rangle\propto r\dot{d}/(D+r^2/2R_\ast)$. The latter expression reveals the radial extent of the hydrodynamic field $a_{h}=(2R_\ast D)^{1/2}$~\cite{Davis1986}: the radial velocity vanishes at $r=0$ and $r\gg a_h$, and is maximal and of the order of $\langle v_r \rangle \propto a_h\dot{d}/D$ at $r\propto a_h$. The laminar assumption implies that $v_r$ is parabolic along the normal $z$ axis~\cite{Leroy2011} ($v_r=0$ at the solid/fluid interfaces due to the non-slip condition and is maximal at $z=0$ for symmetry reason). The local shear rate in the fluid is thus $\dot{\gamma}=\partial v_r/\partial z\propto a_h\dot{d}/D^2$ and its mean value over the whole interstitial region is $\langle\dot{\gamma}\rangle=(1/\pi R^2)\int_0^R{\dot{\gamma}2\pi rdr}\propto(a_h/R)^2\dot{\gamma}\propto\dot{d}/a_h$. The flow generates a hydrodynamic pressure given by the Stokes equation, $(\partial p_h/\partial r\propto p_h/a_h)\simeq(\mu\partial\dot{\gamma}/\partial z\propto \mu a_h\dot{d}/D^3)$~\cite{Davis1986,Leroy2011}, such that $p_h\propto\mu a_h^2\dot{d}/D^3\propto\mu\omega R_\ast d/D^2$. In the case of rigid particles, the interstitial flow consequently induces the well-known Reynolds force~\cite{Guyon2001}, which counteracts the relative approach of the particles, $F_h\simeq\pi a_h^2p_h\propto \mu\omega R_\ast^2 d/D$.\\ 

{\em Interstitial fluid flow}. The details of the interparticles flow regime can be inferred from the long-wavelength experimental data shown in Figs.~\ref{fig:signals_low_P},~\ref{fig:signals_high_P}, and~\ref{fig:wave_speed}. The largest acceleration is typically $\Gamma\propto\omega V\sim10$~m/s$^2$, see Figs.~\ref{fig:signals_low_P} and~\ref{fig:signals_high_P}, with $V$ denoting the velocity field. The rise duration revealing a frequency content at around $f\sim1$~kHz, see Fig.~\ref{fig:wave_speed}, then $V\sim1.6$~mm/s at the most. The normal relative velocity between two particles, $v_n=\dot{d}$, can be deduced from an estimation of the gradient of the velocity field, $V/\lambda$ where $\lambda=c/f$ is the wavelength, as $v_n\propto RV/\lambda\propto R\Gamma/c$. The wave speed being at least $c\sim200$~m/s with fluids, see Fig.~\ref{fig:wave_speed}, then $\dot{d}\sim0.25$~mm/s at the most. This allows estimating then the typical fluid's shear rate in the interstitial region, $\langle\dot{\gamma}\rangle\propto\dot{d}/a_h\propto (R/D)^{1/2}(\Gamma/c)$. The latter requires an estimation of the fluid's thickness $D$. As a worst situation, the surface roughness of the particles, $Ra\sim10$~nm~\cite{Brunet2008,Buttinoni2017}, can be considered as the minimal achievable separation, $D\sim Ra$, below which the fluid may be trapped in between asperities~\cite{Marshall2011,Buttinoni2017}. At worst, the typical shear rate is thus $\langle\dot{\gamma}\rangle\sim25$~s$^{-1}$. This value remains an order of magnitude, or smaller, below the critical shear rate, $\langle\dot{\gamma}\rangle\ll\dot{\gamma}_f=G/\mu$, above which the fluid becomes non-Newtonian: with $G\simeq5$~kPa~\cite{Oswald2014}, the critical shear rate is $\dot{\gamma}_f\sim166$~s$^{-1}$ at the highest viscosity $\mu=30$~Pa.s and $\dot{\gamma}_f\sim50.000$~s$^{-1}$ at the lowest viscosity $\mu=0.1$~Pa.s. The fluid thus remains Newtonian in all our experiments. Finally, the largest radial velocity of the fluid, $\langle v_r \rangle \propto (R/D)^{1/2}(\Gamma R/c)\sim6.25$~cm/s, indicates a Mach number $M_f=\langle v_r \rangle/c_f\ll1$ and a Reynolds number, $Re=\rho_f \langle v_r \rangle D/\mu\ll1$ well below unity. The interstitial flow thus remains incompressible and laminar~\cite{Guyon2001}, in agreement with previous assumptions. As a consequence, the increase of the effective elasticity observed in wet samples cannot be attributed either to a non-Newtonian viscoelastic behavior of the fluid, or to an effect of its compressibility.\\

\begin{figure*}
\includegraphics[width=0.882\textwidth]{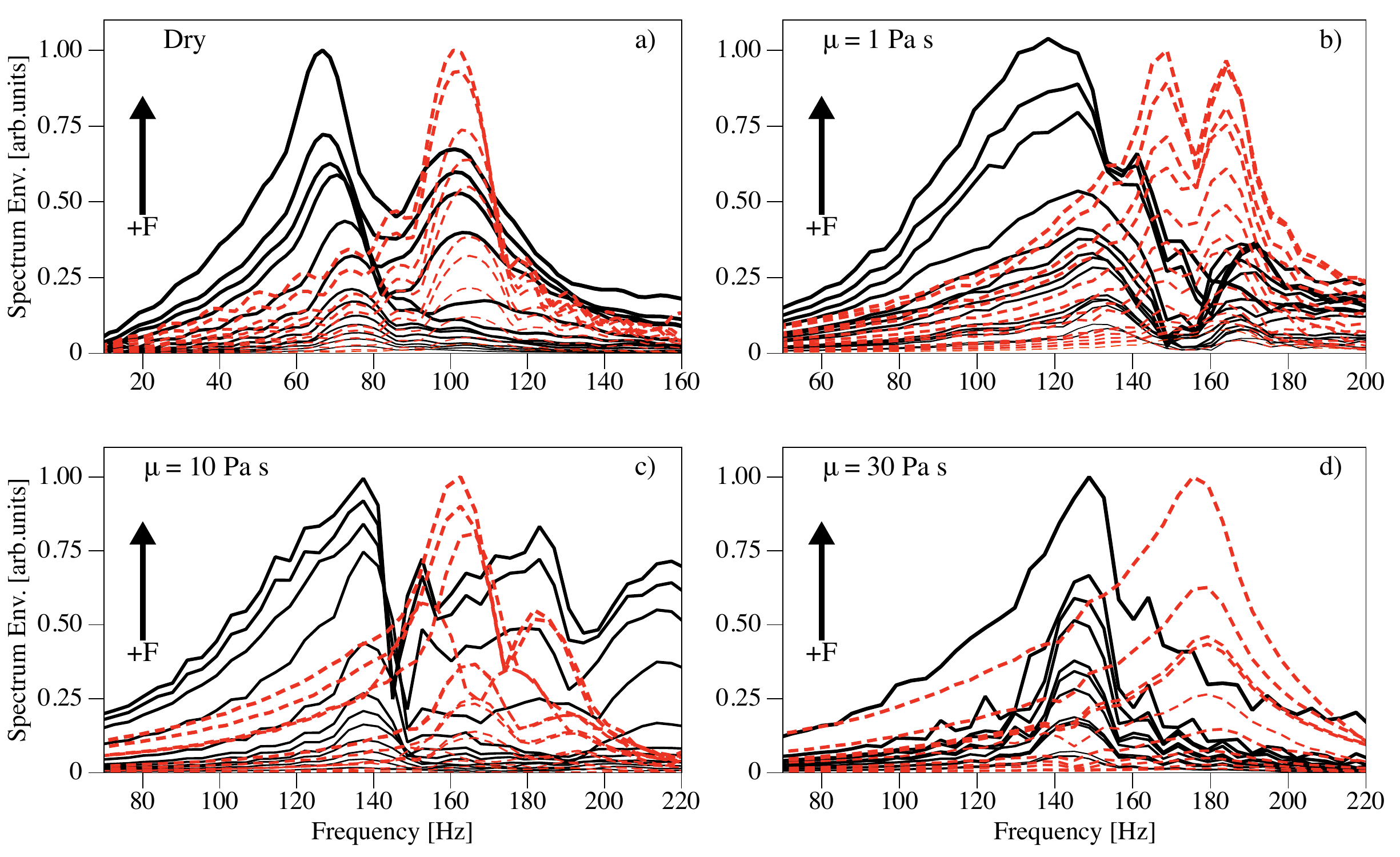}
\caption{\label{fig:weakening_spectra} Frequency spectra at $P_0=3.2$~kPa (solid black line) and $P_0=83$~kPa (dashed red line) for increasing amplitudes of excitation. (a) dry contacts, (b) $\mu=1$~Pa.s, (c) $\mu=10$~Pa.s, and (d) $\mu=30$~Pa.s show the weakening behavior of the lowest mode. The vertical arrows are guides for the eye, pointing toward increasing amplitudes of excitation.}
\end{figure*}

{\em Elastic particles ($\delta_n>0$)}. Nevertheless, the hydrodynamic pressure field $p_h\propto\mu\omega R_\ast d/D^2$ can reach sufficiently high values, for instance when the fluid's thickness vanishes, to involve the elasticity of the particles~\cite{Davis1986,Leroy2011}. Given the Hooke's law, $p_h\propto E_\ast(\delta_n/a_h)$, the elastic deformation of the particles can be rewritten as a fraction of their relative displacement, $(\delta_n/d)\propto (D_c/D)^{3/2}$, where~\cite{Leroy2011}
\begin{equation}
D_c=8R_\ast(\mu\omega/E_\ast)^{2/3}
\label{eq:model_D_cutoff}
\end{equation}
is a cutoff thickness at which the deformation accommodates the displacement, $\delta_n=d$. This situation was referred to as an {\em elastic confinement} of the fluid~\cite{Villey2013}: the fluid being clamped by its viscosity, it does not flow but instead mediates the elastic deformations of the bodies as a rigid layer. The cutoff thickness thus stands as a minimal achievable fluid thickness. Quantitatively, it is approximately $D_c\simeq70$~nm with the lowest viscosity and $D_c\simeq3$~$\mu$m with the highest viscosity, under our experimental conditions. These values are larger than the typical surface roughness, $D_c\gg Ra\sim10$~nm, and smaller than the typical thickness of the liquid coating, $D_c\ll D_{coat}\sim150$~$\mu$m, this regime is thus accessible in our experiments. In such an elastohydrodynamic regime, the typical extent of the field and the normal stiffness thus become, respectively
\begin{eqnarray}
a_{ehd} &=& (2R_\ast D_c)^{1/2}\propto R_\ast(\mu\omega/E_\ast)^{1/3}, \label{eq:model_Aehd}\\
\kappa_n &\propto& F_{ehd}/\delta_n\propto E_{\ast}a_{ehd}, \label{eq:model_Kehd}
\end{eqnarray}
where the elastohydrodynamic force $F_{ehd}\propto\pi a_{ehd}^2p_{ehd}$ is proportional to the pressure given by the Hooke's law, $p_{ehd}\propto E_\ast(\delta_n/a_{ehd})$, with $p_{ehd}\propto \mu\omega R_\ast d/D_c^2$. In addition, the tangential interaction presumably becomes frictionless in presence of the lubricating layer of fluid, $\kappa_t=0$. Consequently, the effective longitudinal modulus given by Eqs.~\ref{eq:model_Kemt},~\ref{eq:model_Gemt},~\ref{eq:model_Aehd}, and~\ref{eq:model_Kehd} becomes
\begin{equation}
(M_{ehd}/E_{\ast}) = \alpha_{ehd} (\mu\omega/E_{\ast})^{1/3},\label{eq:model_Mehd}
\end{equation}
where $\alpha_{ehd}$ is a numerical prefactor depending on $Z$ and $\phi_s$ only. Interestingly, the comparison of the expressions of the dry and the wet elastic moduli, given by Eqs.~\ref{eq:model_Mdry} and~\ref{eq:model_Mehd}, respectively, shows that their ratio is a function of a single nondimensional parameter, $(M_{ehd}/M_{dry})\propto(\mu\omega/p)^{1/3}$. This result substantiates the observation of a nontrivial competition between a viscous contribution of the fluid and the elastic deformation of the particles under the action of the confinement pressure. This feature is confirmed by representing the experimental wet-to-dry elastic moduli as a function of $\mu\omega/p$, see Fig.~\ref{fig:elastic_moduli}(b). The data set, aggregating two different confinement pressures and four different viscosities, indeed fairly collapses along a master curve when represented in such a nondimensional form. However, the ratio of the elastic moduli asymptotically saturates at one for small values of $\mu\omega/p$ in experiments. This is consistent with the fact that a wet sample under a high pressure and with a small interstitial viscosity should tend to behave as a dry sample, $M_{wet}\simeq M_{dry}$ at $\mu\omega/p\ll1$. Moreover, within a constant confinement pressure only, the fluid quasistatically flows out from the contact between particles, leaving a central region in mechanical contact surrounded by a peripheral region filled with fluid~\cite{Marshall2011}, see Fig.~\ref{fig:sketch_contact}. A more convenient ansatz would thus correspond to a Hertzian elastic response, coming from the central and flat dry region, acting in parallel to a peripheral elastohydrodynamic response, due to the fluid pinched at the edge between the elastic solids:
\begin{equation}
M_{wet} = M_{dry}+M_{ehd}.\label{eq:model_Mwet}
\end{equation}

The ansatz given by Eq.~\ref{eq:model_Mwet} is probed in Fig.~\ref{fig:elastic_moduli}(c), which represents the ratio $(M_{ehd}/M_{dry})=(M_{wet}/M_{dry}-1)$ as a function of $(\mu\omega/p)$. As in Fig.~\ref{fig:elastic_moduli}(b), the data set still demonstrates fair correlations, but now shows a clear power law with an exponent close to $1/3$, in agreement with our analysis, see Eqs.~\ref{eq:model_Mdry} and~\ref{eq:model_Mehd}. Quantitatively, the fit of the data shown in Fig.~\ref{fig:elastic_moduli}(c) provides $(\alpha_{ehd}/\alpha_{dry})=1.01\pm43\%$ at low confinement pressure and $(\alpha_{ehd}/\alpha_{dry})=0.46\pm20\%$ at high confinement pressure; this corresponds to prefactors of the order of unity, $\alpha_{ehd}\simeq0.25$ and $\alpha_{ehd}\simeq0.40$, respectively. Finally, the Fig.~\ref{fig:elastic_moduli}(d) shows $(M_{ehd}/E_{\ast})$ as a function of $(\mu\omega/E_{\ast})^{1/3}$ for the high confinement pressure data set. The plot confirms that $M_{ehd}$ increases monotonically according to Eq.~\ref{eq:model_Mehd}, i.e., the effective elastic modulus increases with the viscosity of the fluid; matching the curve to Eq.~\ref{eq:model_Mehd} provides a consistent value of the prefactor, $\alpha_{ehd}=0.37\pm37\%$.\\

The analysis of the experimental wave speed presented in Fig.~\ref{fig:wave_speed}, based on an effective medium theory, thus demonstrates that the propagation of mechanical waves in wet granular samples induces an elastohydrodynamic mechanism at the interparticle level. The mechanical response of a wet sample results from the competition between (i) an elastic contribution related to the static confinement pressure within the contact region between grains and (ii) an elastohydrodynamic interplay between the particles and the fluid, which resides at the periphery of the contacts. The crossover between these two contributions is fairly described by a unique nondimensional number, $(\mu\omega/p)$.

\section{\label{sec:weakening} Material weakening}

Material weakening has been proposed as a triggering mechanism for the emission of secondary pulses after the passage of a principal mechanical event. Several authors~\cite{Johnson2005,Jia2011,Johnson2016} have linked the breaking of unconsolidated and weak contacts with the loss of material strength and the dynamical change of the shape of the {\em coda wave}. Here, the material softening is probed by tracking the frequency shift of the lowest prominent mode in the spectrum of the long-lasting outgoing acceleration. The dry configuration is first considered, see Fig.~\ref{fig:weakening_spectra}(a), at both low and high confinement pressure. It is observed that for the high confinement pressure, the low-frequency component, at around $100$~Hz, remains unaffected by the strength of the dynamical perturbation. This results is coherent with the fact that the dynamics of the sample is linear at high confinement, owing to a negligible dynamical perturbation: the sample is highly consolidated and no material weakening is observable. On the other hand, at the lowest confinement pressure, the low-frequency component decreases with increasing impulse amplitude: a $20\%$ variation in respect to the initial frequency is observed, consistently with the observations available in the literature~\cite{Johnson2005}. The decrease of the prominent frequency reveals the nonlinear nature of the contact dynamics and the weakening of the material as the dynamical strength is progressively increased.\\

\begin{figure}[t]
\includegraphics[width=0.450\textwidth]{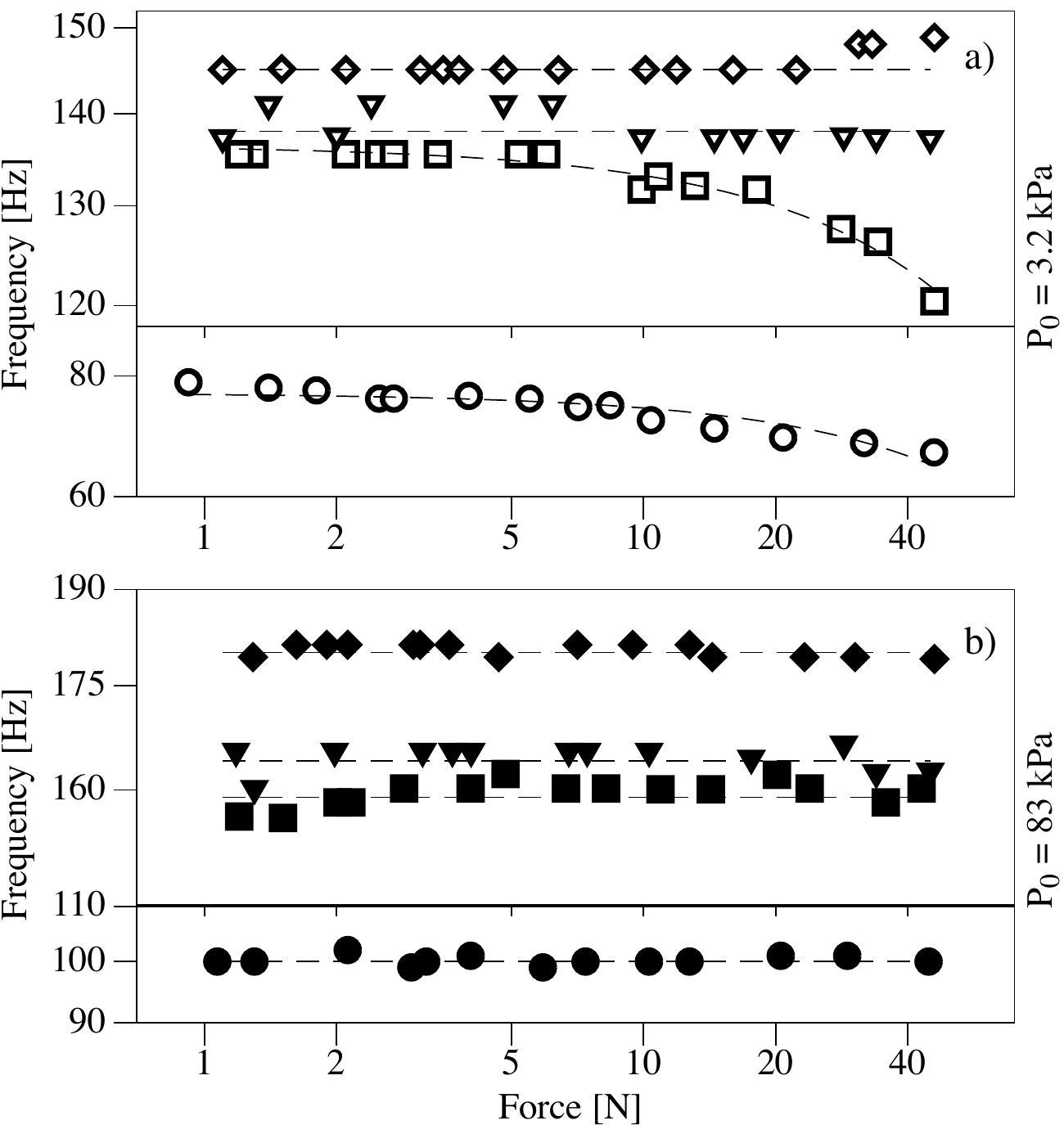}
\caption{\label{fig:weakening_frequency} Frequency of the lowest mode as a function of the input force, for different fluid viscosities at (a) low and (b) high confinement pressures. The markers refer to the definitions given in Fig.~\ref{fig:wave_speed}.}
\end{figure}

Next, the same procedure is repeated in samples containing interstitial fluids with different viscosities, see Figs.~\ref{fig:weakening_spectra}(b)--\ref{fig:weakening_spectra}(d). In Fig.~\ref{fig:weakening_frequency}, the evolution of the lowest mode for every fluid is shown as a function of the impulse amplitude. The weakening of the sample remains observable for the lowest viscosity only, and progressively disappears when the viscosity is increased. This means that an interstitial fluid with a sufficiently high viscosity consolidates the sample and tends to linearize its mechanical response. Indeed, using the nondimensional parameter $(\mu\omega/p)$ derived in Sec.~\ref{sec:model} as an indicator of the regime, one find a crossover viscosity $\mu_{\ast}=p/\omega$. Below this value, the dynamics is essentially Hertzian and nonlinear, see Eqs.~\ref{eq:model_Mdry}. Above the crossover, the dynamics tends to becomes linear owing to a dominant elastohydrodynamic interplay, see Eqs.~\ref{eq:model_Mehd} and~\ref{eq:model_Mwet}. Such a crossover is $\mu_{\ast}\simeq3.6$~Pa.s at low confinement pressure, where $f\sim140$~Hz, in agreement with our observations. Similarly, the crossover is $\mu_{\ast}\simeq82$~Pa.s at high confinement pressure, where $f\sim160$~Hz. In this case, the behavior of all the wet samples relies on the Hertzian elastic interaction: it is thus reminiscent to the dry configuration, which proved to be consolidated and where no weakening is expected.\\

\section{\label{sec:conclusion} Conclusions}

The experimental evidences presented in this paper showed that the presence of an interstitial fluid in a granular media significantly modifies the contact dynamics between particles and, consequently, the features of mechanical waves propagation in the long wavelength approximation. In both low and high confinement cases, it was observed that the fluid induces an elastohydrodynamic mechanism that enhances the rigidity of the contacts, rendering a higher wave speed as compared with the dry configuration. All our results were discussed in terms of an effective mean-field theory, coupled to a description of the elastohydrodynamic interaction between spherical elastic particles mediated by an interstitial Newtonian viscous fluid. Our analysis suggested a nontrivial competition between the elastohydrodynamic interaction and the elastic deformation due to the confining pressure. The interplay was shown to be fairly described by a unique nondimensional parameter $\mu\omega/p$, which allowed defining a threshold viscosity above which the effect of the fluid dominates. It was also observed that the dry granular material weakens when submitted to strong dynamical perturbations, due to the breaking of unconsolidated contacts. This nonlinear softening can be impeded by either increasing the confinement pressure, or by adding an interstitial fluid with a viscosity above the threshold. Our analysis and description qualitatively match the experimental observations; all these results might prove to be useful in practical situations and pave the way to the need of a more quantitative and precise description.\\

\begin{acknowledgments}
The authors thank Francisco Melo for many fruitful discussions. R.Z. acknowledges CONICYT National Doctoral Program Grant No. 21161404 and the financial support of {\em Supm{\'e}ca} during his stay in France. S.J. acknowledges the {\em Pontificia Universidad Cat{\'o}lica de Valpara{\'i}so} and the Franco-Chilean {\em Laboratoire International Associ{\'e} LIA-MSD} for the financial support during his stay in Chile. F.S. acknowledges the financial support from FONDECYT Project No. 11140556.
\end{acknowledgments}

\bibliographystyle{unsrt}
\bibliography{1808_03150_v2}

\end{document}